# An Approach Detecting the Event Horizon of Sgr A*

Makoto Miyoshi[1], Seiji Kameno[2],
José K. Ishitsuka[3], Zhi-Qiang Shen[4], Rohta Takahashi[5]
and
Shinji Horiuchi[6]

**Abstract**

Imaging the vicinity of a black hole is one of the ultimate goals of VLBI astronomy. The closest massive black hole, Sgr A*, located at the Galactic center, is the leading candidate for such observations. Combined with recent VLBI recording technique and submillimeter radio engineering, we now have sufficient sensitivity for the observations. Here we show performance simulations of submillimeter VLBI arrays for imaging Sgr A*. Good images are obtained from submillimeter VLBI arrays in the southern hemisphere composed of more than 10 stations. We also note that even with a small array, we can estimate the shadow size and then the mass of the black hole from visibility analysis. Now, all we need is to construct a submillimeter VLBI array in the southern hemisphere if we wish to unveil the black hole environment of Sgr A*.

## 1. Introduction

Imaging a black hole system is one of the final goals in VLBI astronomy. Sgr A*, the massive black hole at the Galactic center, is the leading candidate for the research. This is not merely because Sgr A* is the most convincing black hole candidate (Schödel et al. 2002, Ghez et al. 2000, Shen et al. 2005) but mainly because Sgr A* has the event horizon with the largest apparent angular size among black hole candidates. In table 1 we show the apparent angular sizes of Schwarzschild radii of several black hole candidates. The Sgr A* has the largest apparent angular Schwarzschild radius that is estimated to be $6\mu$arcseconds from the mass ($2.6 \times 10^6 M_\odot$, Ghez et al. 2000) and the distance of the Galactic center (8 kpc).

As Schwarzschild radius ($R_s$) is proportional to the mass of the object ($R_s = 2GM_{BH}/c^2$, where $G$ is gravitational constant, $M_{BH}$ is the mass of black hole, and $c$ is the light velocity), the real Schwarzschild radii of what you call super massive black holes with mass more than a few $\times 10^7 M_\odot$ are really large. However, such massive black holes are located at very great distances more than a few Mpc, then their resultant apparent angular sizes are not so large.

As for stellar black holes in our Galaxy located closer to us, because the masses are only a few $M_\odot$, the apparent angular Schwarzschild radii are quite small. For instance, a stellar black hole with $1 M_\odot$ located at 1 pc has the apparent angular Schwarzschild radius of 0.02 $\mu$arcseconds.

The mass of Sgr A* is a few of $10^6 M_\odot$, which is not so large as those of other massive black holes. But the distance of Sgr A* is 8 kpc at most, three orders of magnitude closer than any other massive black hole. Then Sgr A* shows the largest apparent angular Schwarzschild radius of all black hole candidates (See table 1).

Needless to say, we cannot see a black hole itself alone in dark space. However we can expect to see the shadow of a black hole like a silhouette when the black hole is enveloped by the luminous emission of a jet or accreting hot matter. Views of black holes in such situations have been theoretically investigated by many theorists (Cunningham & Bardeen 1972, Bardeen & Cunningham 1973, Cunningham 1975, Lumine 1979, Sikora 1979, Fukue & Yokoyama 1988, Perez & Wagoner 1991, Jaroszynski, Wambsganss, & Paczynski 1992, Chandrasekhar 1983, Kindl 1995, Hollywood & Melia 1995, Quien, Wehrse & Kindl 1996, Hollywood & Melia 1997, Bromley, Miller & Pariev 1998, Pariev & Bromley 1998, Usui, Nishida & Eriguchi 1998, Falcke, Melia, & Agol 2000, Bromley, Melia, & Liu 2001,

---
[1] National Astronomical Observatory, 2-21-1 Osawa, Mitaka, Tokyo 181-8588; makoto.miyoshi@nao.ac.jp
[2] Kagoshima University, 1-21-35 Korimoto, Kagoshima 890-0065, Japan; kameno@sci.kagoshima-u.ac.jp
[3] Calle Badajos 169, Urb. Mayorazo, Cuarta Etapa Ate Vitarte, Postal Code 03, Lima, Peru; pepe@geo.igp.gob.pe
[4] Shanghai Astronomical Observatory, Chinese Academy of Sciences, Shanghai 200030, China; zshen@shao.ac.cn
[5] Department of Earth Science and Astronomy, Graduate School of Arts and Science, University of Tokyo, Komaba, Meguro, Tokyo 153-8902, Japan; rohta@provence.c.u-tokyo.ac.jp
[6] Center for Astrophysics and Supercomputing, Swinburne University of Technology, Mail stops 31 P. O. Box 218 Hawthorn, Victoria, 3122, Australia; shoriuchi@swin.edu.au

Fukue 2003, Takahashi 2004). The typical diameter of a black hole shadow is around 5 Schwarzschild radii.

The finding of a black hole shadow, namely the observation of an event horizon, is the perfect evidence of black hole. The apparent angular size of the black hole shadow of Sgr A* is about 30 $\mu$arcseconds in diameter. Recent observations indicate the mass of Sgr A* is $3.7 - 4.1 \times 10^6 M_\odot$ (Schödel et al. 2002, Ghez et al. 2003). If we accept the newly estimated mass, the size of the black hole shadow of Sgr A* is more than 45 $\mu$arcseconds in diameter.

Sgr A* was detected 30 years ago (Balick & Brown 1974) and has long been recognized to be a very quiet and stable source. In recent years, however, after detection of a radio variation of 106 days in Sgr A* (Zhao et al. 2001), several short time flaring events of Sgr A* have been unveiled. The detected rapid flares of Sgr A* range from a few hours to 30 min at radio, infrared, and x-ray emissions (Miyazaki et al. 2003, Zhao et al. 2004, Genzel et al. 2003, Baganoff et al. 2001, Goldwurm et al. 2003, Porquet et al. 2003). These rapid changes strongly suggest structural change of the accretion disk or eruption of a jet in Sgr A*. Sgr A* has become very important for investigating black hole environments.

Not a few VLBI observations have been performed to unveil the features of Sgr A*. However, the scattering effects by surrounding plasma have blurred the intrinsic radio image and previous VLBI observations at frequencies lower than 86 GHz have been unable to reach the true face so far (Doeleman et al. 2001, Zensus et al. 1999, Bower et al. 1998, Yusef-Zadeh et al. 1994, Rogers et al. 1994, Alberdi et al. 1993, Krichbaum et al. 1993, Marcaide et al. 1992, Jauncey et al. 1989, Lo et al. 1985, 1998, Bower et al. 2004, Shen et al. 2005).

Because the scattering effect is proportional to the square of the observing wavelength, the effects become negligible at submillimeter observations. To unveil the intrinsic image, we should accomplish submillimeter VLBI observations of Sgr A* (Falcke et al. 2000)

As the first step of planning such a submillimeter VLBI array, we show simulation results of performance of submillimeter VLBI array configurations for Sgr A* observations.

We also point out the capability of visibility analysis obtained from VLBI array composed of few stations.

## 2. Simulations

We performed simulations for testing array performance whether the black hole shadow of Sgr A* can be recognized or not. Because Sgr A* is located at –30° in declination, suitable arrays will be located in the southern hemisphere. We checked performance of three virtual arrays in the southern hemisphere, the VLBA configuration, a realistic network connecting submillimeter interferometers and so on. As image models of black hole shadow in Sgr A* we use two kinds, one a black hole shadow embedded at Gaussian brightness distribution, the other a black hole shadow in the accreting disk viewed from an edge-on angle.

As we focus on the performance of array configurations, the sensitivity of every station is unified. Namely the antenna diameter is 12 m with an aperture efficiency of 0.7. The system temperature at 230 GHz is 150 K which will be attainable at the ALMA site. The observing bandwidth is 1000 MHz. Atmospheric conditions are essential in real submillimeter observations on the ground, but are neglected

Table 1. Black Hole Mass, Distance & Shadow Size.

| Object | Mass [$M_\odot$] | Distance [kpc] | Schwarzschild Radius [km] | Schwarzschild Radius [A.U.] | [$\mu$arcsec] | Shadow (diameter) [$\mu$arcsec] |
|---|---|---|---|---|---|---|
| (1) | (2) | (3) | (4) | (5) | (6) | (7) |
| a stellar black hole | 1 | $1 \times 10^{-3}$ | $2.95 \times 10^3$ | $1.97 \times 10^{-8}$ | 0.02 | 0.10 |
| M82 | $1.0 \times 10^{3(a)}$ | 3700 | $2.95 \times 10^9$ | $1.97 \times 10^{-2}$ | 0.01 | 0.05 |
| Sgr A* | $2.6 \times 10^{6(b)}$ | 8 | $7.67 \times 10^9$ | $5.11 \times 10^{-2}$ | 6.39 | 31.96 |
| Sgr A* | $4.1 \times 10^{6(c)}$ | 8 | $1.09 \times 10^{10}$ | $7.28 \times 10^{-2}$ | 9.10 | 45.48 |
| M31 | $3.5 \times 10^{7(d)}$ | 800 | $1.03 \times 10^{11}$ | $6.88 \times 10^{-1}$ | 0.86 | 4.30 |
| NGC4258 (M106) | $3.9 \times 10^{7(e)}$ | 7200 | $1.15 \times 10^{11}$ | $7.76 \times 10^{-1}$ | 0.11 | 0.53 |
| M87 | $3.2 \times 10^{9(f)}$ | 16100 | $9.44 \times 10^{12}$ | $6.29 \times 10^{+1}$ | 3.91 | 19.54 |

Notes. – Col. (1): Object name. Col. (2): Estimated mass of black hole. a) Matsumoto & Tsuru (1999), Ptak & Griffiths (1999), b) Ghez et al. (2000), c) Ghez et al. (2003), d) Kormendy & Bender (1999) e) Miyoshi et al. (1995), Herrnstein et al. (1999), f) Ford et al. (1994) Col. (3): Object distance. Cols. (4), (5) and (6): real Schwarzschild radius in km and in A.U., apparent angular size in $\mu$arcsec respectively. Col. (7): Apparent angular diameter ($\mu$arcsec) of black hole shadow ($=5 \times R_s$).

here. We will show such simulations using the ARIS package (Asaki et. al 2007) in our next paper.

## 2.1. Array configurations

Here we select the following 8 array configurations for the simulations.

- Array A: the same location as that of the VLBA (NRAO). Needless to say, the actual VLBA antennas have neither 230 GHz receivers nor sufficient antenna surface accuracy. This is only for configuration simulations.
- Array B: the VLBA configuration plus a virtual station located at Huancayo in Peru. The position of the virtual Huancayo antenna is situated 3375 m in altitude, just where latitude 12.0° S meets longitude 75.3°W.
- Array C: the VLBA configuration plus the Huancayo station and the ALMA in Chile which is at longitude 67.4° W by latitude 23.0° S, 5000-m in altitude.
- Array D: the VLBA plus the Huancayo station, the ALMA and the SEST (ESO) in Chile. The location of the SEST is at longitude 70.7° W by latitude 29.3° S, 2400-m in altitude. (NOTE: The SEST closed its operations several years ago.)
- Array E: this array includes realistic submillimeter interferometers: namely the SMA at Mauna Kea in Hawaii, the CARMA in eastern California, the virtual Huancayo, the ALMA and the SEST (ESO) in Chile. The SMA and the CARMA are now in operation, while the ALMA is under construction.
- Array F: the inversed VLBA, located in the southern hemisphere. Except the latitudes of stations, all other parameters are the same as those of Array A.
- Array G: a virtual array located in the southern hemisphere. This array is composed of 9 stations in South America and one station at the SAAO in South Africa. The locations are listed in table 2. Except Itapetinga in Brazil, the other 8 stations in South America are situated at higher than 2400 m.
- Array H: a virtual array located mainly in the southern hemisphere. This array includes the array G denoted above, the SMA and the CARMA in the northern hemisphere.

Because the atmospheric fluctuations are quite large when the elevation angle of the observing object is below 10°, we limit the uv coverage with the elevation angles in each station higher than 10°.

Table 2 Array G: station positions.

| Station | Latitude | Longitude | Altitude |
|---|---|---|---|
| Huancayo | –12.0° | 75.3° | 3375 m |
| ALMA | –23.0° | 67.4° | 5000 m |
| SEST | –29.3° | 70.7° | 2400 m |
| Itapetinga | –23.2° | 46.6° | 800 m |
| SAAO | –32.4° | 20.8° | 1760 m |
| Cerro Murallon | –49.8° | 73.5° | 3600 m |
| Cotopaxi | –1.0° | 77.0° | 5896 m |
| Pico Cristobal | 11.0° | 74.0° | 5684 m |
| Maipo | –34.0° | 71.0° | 5290 m |
| Araral | –21.5° | 67.6° | 5680 m |

Figure 1 shows the uv coverage of the 8 arrays mentioned above (in case of observing frequency, 230 GHz). The uv coverage of VLBA (Array A) is notoriously worse in a north-south direction, deficient for imaging Sgr A* (fig1. a) (Bower et al 1999). Additions of stations in South America reinforce the north-south coverage of the VLBA alone (fig.1. b, c, d). The uv coverage of Array E (a realistic submillimeter VLBI array) is very wide but quite sparse for Sgr A* (fig. 1 e).

The corresponding synthesized beams (or dirty beams) are shown in figure 2. Also in table 2 we show each restoring beam size (Gaussian shape) obtained by fitting to the dirty beam in the task IMAGR of AIPS. While all of the FWHM of minor axes are comparable with or smaller than the diameter of the black hole shadow of Sgr A*, all the FWHM of major axes are larger.

Table 3. Restoring normal beam sizes calculated in AIPS.

| Array | Location | FWHM of Major and Minor Axes | Position Angle of Major Axis |
|---|---|---|---|
| A | VLBA anntenae location | 76.833 × 26.115 μasec | –2.1° |
| B | VLBA plus Huancayo in Peru | 42.317 × 24.207 μasec | 24.9° |
| C | VLBA plus Huancayo, ALMA | 40.359 × 22.234 μasec | 41.4° |
| D | VLBA plus Huancayo, ALMA, SEST | 39.513 × 20.905 μasec | 48.2° |
| E | SMA, CARMA, Huancayo, ALMA & SEST | 52.162 × 15.582 μasec | 42.6° |
| F | inversed VLBA location in latitude | 45.979 × 31.480 μasec | 5.7° |
| G | virtual 9 (South America) plus SAAO | 45.456 × 31.430 μasec | 64.2° |
| H | array of the G plus SMA, CARMA | 40.679 × 19.425 μasec | 47.7° |

The synthesized main beams (spatial resolutions) of the array B – H are about three times smaller in declination than that of array A as shown in figure 2. The main beam of array E is certainly small but also has quite high-level side lobes comparable to that of the main beam. The arrays F, G, and H show high main beams and quite low side lobe levels.

### 2.2. Image models for Sgr A*

Here we use two image models for Sgr A*. One is a Gaussian shape with central black hole shadow (fig.3, M). From the first VLBI observations at 215 GHz the outer size of Sgr A* is estimated to be about 0.1 mas in diameter (Krichbaum et al. 1998). We adopted the value as the outer diameter of image model A. We use here the previous estimated mass of $2.6 \times 10^6 M_\odot$ (Ghez et al. 2000) and the corresponding shadow size of $30\mu$arcseconds in diameter. The shape of the Gaussian brightness distribution is with a major axis of 0.1 mas (FWHM) and minor axis of 0.08 mas (FWHM). The position angle (PA) of the major axis is 80°. This shape is after previous VLBI observations at a lower frequency (43 GHz) that show east-west elongation of the apparent shape of Sgr A*. Generally the elongation has not been interpreted as any kind of intrinsic structure, but the effect of anisotropic scattering, and the shape adopted here is only for performance tests of arrays. The central shadow shape is also the same elliptical shape with $30\mu \times 24\mu$ arc seconds, and PA=80°.

The other image model B is a type of accretion disk viewed from an edge-on angle plus very faint halo (figure 4 (M)). This model is produced from numerical ray tracing by Takahashi (2004). The viewing angle to the disk plane is 89°. The outer diameter of the disk is $40R_s$, or $240\mu$as in apparent angular diameter. The spin parameter of the black hole is zero, namely this is a Schwarzschild black hole. The image shows a quite complex figure strongly affected by gravitational lensing and by Doppler boosting with relativistic motion of the disk. To the left of the black hole shadow, appears the brightest area which is caused by Doppler boosting to our line of sight, while on the right, the shadow seems to be slightly elongated because Doppler de-boosting decreases the brightness of the area of the accretion disk. At the top of the black hole shadow we can see the gravitationally lensed image of the opposite side of the disk. The brightest position caused by Doppler boosting is located about $14\mu$ arc seconds east (left) of the center. The brightness of the point of symmetry is about 5.5% of that of the peak. The brightness of the faint halo is proportional to the inversed square of the distance from the center. In Figure 4 (M) the halo shows with the $10^{-4}$ and $10^{-3}$ levels of the peak brightness by contours. The brightness ratio between the maximum to the dark halo is about 700 times.

The adopted flux density of the image models is 3 Jy which is the typical flux density of Sgr A* at 230 GHz.

The second image model B is really fantastic and shows typical physical phenomena we expect in black hole vicinities. However, we suppose the real image of the black hole shadow of Sgr A* should be similar to the first image model A. We will discuss the issue later.

The exact frequency free from scattering effect is not sure. For example, Falcke et al. (2000) assumed the frequency is 500 GHz. Recent millimeter-wave VLBI observations at 43 and 86 GHz show that the scattering effect on Sgr A* radio image weakens, or at least observed image size deviates from the Lambda square-law of the scattering (Shen et. al 2005, Bower et al. 2004). Anyway it is difficult to estimate exactly at which frequency the effect becomes negligible. From the point of view of atmospheric conditions, the observable highest frequency of ground-based VLBI will be 340 GHz at most. This frequency is lower than those observable by single dish telescopes and connected compact interferometers. In order for high-frequency VLBI observations to succeed, weather conditions of most stations should be fine at the same time, which is the main factor limiting high-frequency VLBI performance. We here simulate with the 230 GHz case.

### 2.3. Resultant Images from Clean Deconvolutions

We used the AIPS (NRAO) for our simulations. We add the appropriate thermal noise for the system sensitivities when the faked visibilities are produced from the image models using UVCON in AIPS. Clean deconvolved images were produced with the task IMAGR. The restoring beams are unified to the circular Gaussian with the FWHM of $20\mu$ arcseconds which is smaller than those of the normal restoring beams shown in table 2. From imaging simulations and practical experiences we found usage of somewhat smaller restoring beam (super resolutions) is valid to get higher resolution imagings (for example see the map of Sgr A* with super resolution in Shen et al. 2005). The criteria for judging if the array configuration performance is good or bad is wheter a dark area will appear or not at the corresponding position of the black hole shadow in the model image. In addition it is much better when the resultant images reproduce other fine structure of the image models.

Figure 3 shows the resultant images for image model A. The images produced from arrays A and E are not so good, while other images have a dark area at the center. Array F (inverse VLBA), G and H show good images of black hole shadow. All of these arrays extend to more than 6 giga-wavelengths, and are composed of more than 10 stations mostly located in the southern hemisphere.

Array E, a network of realistic submillimeter array composed of 5 stations – SMA, CARMA, SEST, Huancayo, and ALMA – is insufficient for imaging the black hole shadow in image model A.

Figure 4 shows the resultant images for image model B (the edge-on disk model). Every result clearly shows the gravitationally lensed feature of the opposite side of the disk and the Doppler boosted side of the disk (= the left side of the disk). The right side of the disk that is Doppler de-boosted is also vaguely grasped. As a result the black hole shadow at the center is clearly recognizable. However, the thin disk portion near our sight could not be reproduced in these results. The halo area is also recognizable in every result. In the case of F, G, and H, the circular profile of the halo is clearly shown. Image model B has a complex structure but is reproduced to some degree because the whole size is about two times larger than the size of image model A.

Judging from the simulations the suitable arrays for imaging the Sgr A* black hole shadow require more than 10 stations located in the southern hemisphere extending to 8000 km if the observing frequency is 230 GHz. Addition of stations in the northern hemisphere improves the image.

We also simulated the image quality by changing the array sensitivity and found the effect of sensitivity does not show a larger difference than expected. Rather systemic phase errors from insufficient removal of delay offset and rapid phase change by the atmosphere often damage the images, which is something beyond our scope in this paper but must be seriously considered.

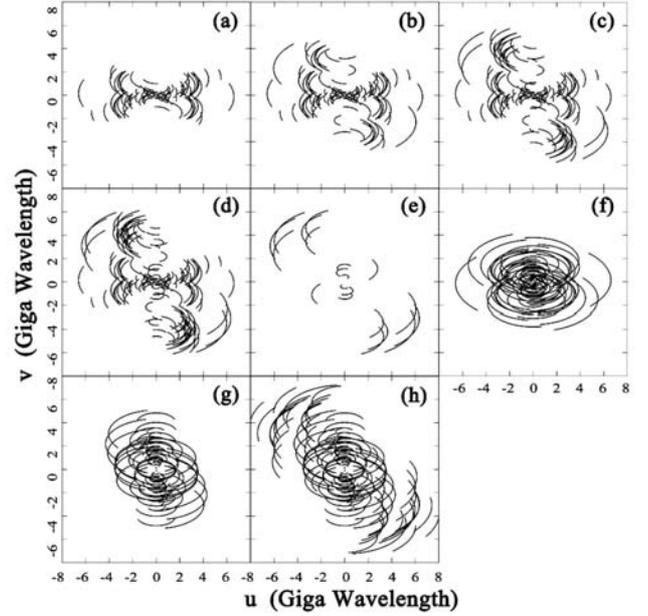

**Figure 1.** The UV coverage plots of the arrays for Sgr A*: (a) Array A, (b) Array B, (c) Array C, (d) Array D, (e) Array E, (f) Array F, (g) Array G and (h) Array H. The span of each side is 16 (−8 to +8) giga-wavelengths (230GHz).

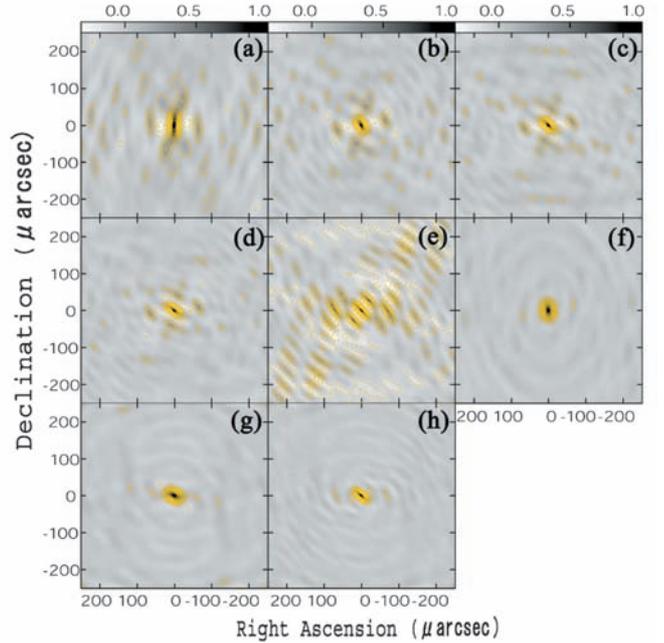

**Figure 2.** The corresponding synthesized (dirty) beams: (a) Array A, (b) Array B, (c) Array C, (d) Array D, (e) Array E, (f) Array F, (g) Array G and (h) Array H. The span of each side is 500$\mu asec$. The step of contours is 10% of the peak brightness.

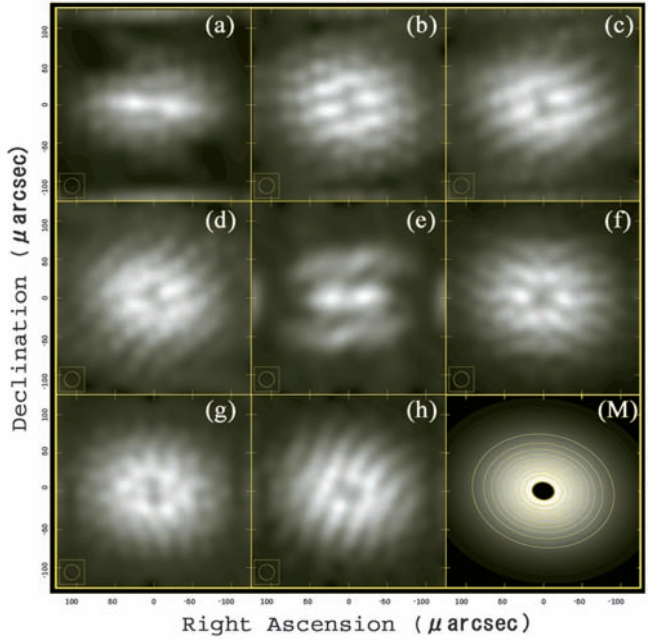

**Figure 3.** Clean results from simulations for image model A (Gaussian brightness distribution with central shadow. (a) Array A, (b) Array B, (c) Array C, (d) Array D, (e) Array E, (f) Array F, (g) Array G and (h) Array H and (M) image model A. The span of each side is 250μasec. The contour levels in image model A (M) are 10% steps. The inset in every panel shows the restoring beam size used.

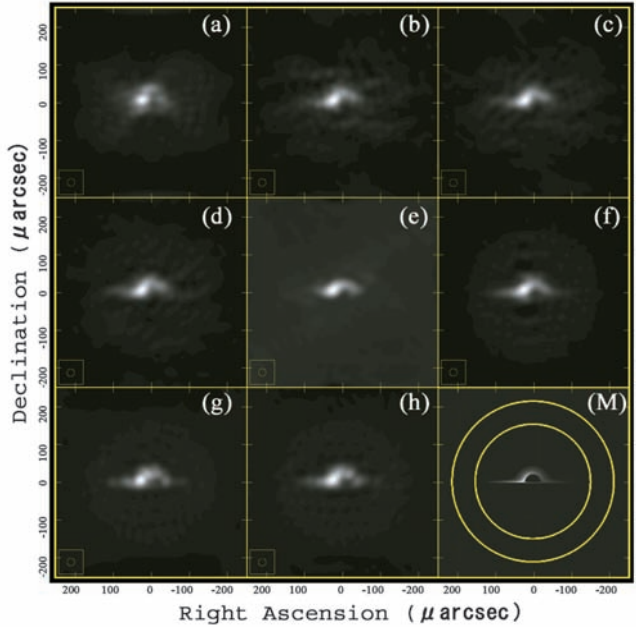

**Figure 4.** Clean results from simulations for image model B (edge-on view of a standard disk plus dark halo). (a) Array A, (b) Array B, (c) Array C, (d) Array D, (e) Array E, (f) Array F, (g) Array G and (h) Array H and (M) image model B. The two contours show 0.01% (outside) and 0.1% (inside) level of the peak brightness. The span of each side is 250μasec. The inset in every panel shows the restoring beam size used.

### 3. Another Method – Visibility Analysis

When an array has limited coverage in the u-v plane, instead of imaging synthesis, visibility analysis has frequently been performed in order to estimate the shape and size of the observed sources. In the early days of radio interferometers such methods were mainly used (e.g. Thompson, Moran, and Swenson, 1986).

Fig. 5 shows visibility amplitude curves of three simple image models, (a) a simple Gaussian brightness without shadow, (b) a Gaussian with a shadow of $30\mu$arcseconds ($M_{BH} = 2.6 \times 10^6 M_\odot$) and (c) a Gaussian with a shadow of $45\mu$arcseconds ($M_{BH} = 3.7 \times 10^6 M_\odot$). For simplicity here we used point-symmetric images.

A Gaussian brightness distribution also shows a Gaussian curve in the visibility amplitudes. If the shadow exists, the visibility function has null value points at some projected baseline length. The null value positions change with the size of the shadow. From the visibility amplitude function, we can distinguish whether the shadow exists or not. Further, because the null value points move according to the shadow size, we can estimate the shadow size, which also means we can measure the mass of the black hole from the null value positions. For measuring the correlated flux densities with uv distance, a small array composed of a few stations is sufficient. Extremely speaking, only one VLBI baseline is sufficient for the purpose.

It is certain that the null value points will appear in visibility amplitude and projected baseline diagrams with other types of images. One of the typical ones is double sources like core & jet structures, which are frequently observed from other AGNs. In case of Sgr A*, there is no convincing result showing separated structures from previous VLBI observations. The VLBI image of Sgr A* always shows a single structure. When an eruption happens at Sgr A*, the double structure will appear. But the structure is a tentative one, and will soon disappear. It will be easy to distinguish the reason for the existence of null value points in visibility amplitude whether from core & jet structures or from the darkness of a black hole shadow. Even if the Sgr A* structure is not simple like the three models used here, we can limit the structure model from estimated mass and spin of the black hole in Sgr A*, and the accretion disk model and shape of black hole shadow from ray-tracing calculations. Reliable image-model fitting to observed visibilities will certainly be possible to detect the black hole shadow in Sgr A*.

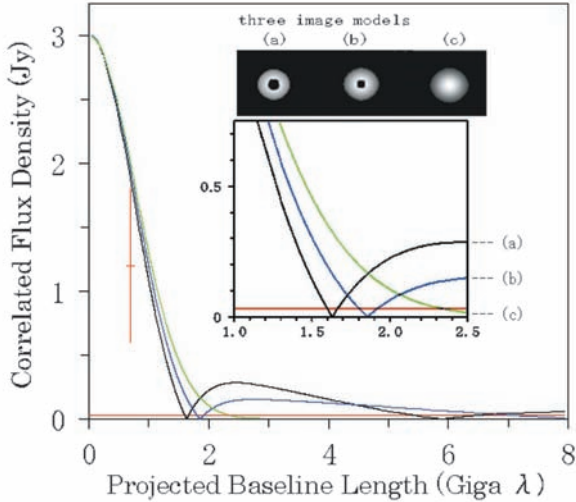

**Figure 5.** The visibility amplitudes of three image models as function of projected baseline: (a) the case of $M_{BH} = 3.7 \times 10^6 M_\odot$, (b) the case of $M_{BH} = 2.6 \times 10^6 M_\odot$ and (c) the case with no black hole or the scattering effect is still dominant. The functions of (a) and (b) have null value points that indicate the existence of the central black shadow. The $3\sigma$ noise level of present engineering performance is shown by the red horizontal line. The red point with error bar is the visibility amplitude measured by Krichbaum et al. (1998).

## 4. Discussions and Conclusions

We used the two types of image models for the simulations. Which image is appropriate for Sgr A*? The spectra of Sgr A* are well fitted to advection-dominated accretion flows (ADAFs, Narayan et al. 1994) or radiatively inefficient accretion flows (e.g. Quataert et al. 1999, Yuan et al. 2003). It suggests that matter density around the black hole in Sgr A* is quite low and then an optically thick disk is improbable (Takahashi 2004). This is not preferable to image model B. In the case of image model B, correlated flux densities at long projected baselines are larger than those of image model A, because image model B includes a very high brightness point caused by Doppler boosting of relativistic velocity of the disk rotation. In other words the observed brightness temperature of Sgr A* should be higher if the image is similar to image model B. Krichbaum et al. (1998) shows that the 215 GHz correlated flux density of Sgr A* at $700 M\lambda$ is about 0.7 Jy, which is consistent with or lower than the predicted flux density from image model A.

If the true image of Sgr A* at submillimeter wavelength is like image model A, from visibility amplitude function with projected baseline length we can easily estimate the diameter of the black hole shadow. The estimated mass of Sgr A* ranges from $2.6 - 4.1 \times 10^6 M_\odot$, which is quite a precise value so that we can safely forecast where the corresponding null points will appear. The first null point appears at the projected baseline ranging $1000 \sim 2000$ km in the case of image model A. If the scattering effect really becomes negligible at 230 GHz, we should put stations for submillimeter VLBI in the Andes mountains at an appropriate distance from the ALMA in order to detect the black hole shadow. The finding of the visibility null points is the first observational evidence of the event horizon of a black hole.

**Table 4.** Parameters related to sensitivity attainable with recent technique.

| name | value |
|---|---|
| antenna diameter | 15 m |
| aperture efficiency $e_\epsilon$ | 0.7 |
| system temperature | 150 K |
| quatized efficiency $e_{q\epsilon}$ | 0.7 |
| integration time | 100 sec |
| bandwidth | 1024 MHz |
| $1\sigma$ noise level | 10 mJy |

We also check the sensitivity for detection of the null points. In figure 4, the horizontal red line shows the $3\sigma$ r. m. s. noise level (30mJy) calculated from the condition in Table 4. The accuracy of the null point measurement is about 100 $M\lambda$ at worst that corresponds to $4\sim 5\mu$ arc seconds in diameter, giving us an accuracy of mass measurement about $3 \sim 4 \times 10^5 M_\odot$. It is obvious that we can distinguish the difference between image model A and B.

The assumed parameters in table 4 are now in our hands using recent VLBI recording technique and submillimeter radio engineering. Last century, the VLBI Giga Bit Recorder attained high-speed recording at 1 Gbps (e.g. Nakajima et al. 1997, Sekido et al. 1999) and much higher recording systems will appear. Even with tandem use of the GBR system 1024 MHz bandwidth recording is attainable. Antenna and receiver systems for submillimeter observations are now making great progress for ALMA use. Phase-up ALMA or ACA will afford great improvement of sensitivity in submillimeter VLBI. The outstanding issue for detecting the event horizon of Sgr A* is whether we have some sites suitable for submillimeter observations in the Andes but away from the ALMA site. Good atmospheric conditions are required. Hence, site survey at the Andes is the key to the project. Now, if only we find suitable sites and construct a submillimeter VLBI array in the southern hemisphere, we can unveil the black hole environments of Sgr A*.

Acknowledgements: We would like to thank all of the attendees of the Japanese VLBI consortium symposium 2003 for helpful discussions. We also thank Dr. Fukue, Dr. Oyama, Dr. Asada, Dr. Nagai and Dr. Doi for their kindness.